\documentclass[final,numberedheadings,sort&compress,numbers]{aipproc}

\usepackage{amssymb}
\usepackage{amsmath}

\newcommand\F{{\scriptscriptstyle\rm F}}
\newcommand\rsL{{\scriptscriptstyle\rm L}}
\newcommand\rsR{{\scriptscriptstyle\rm R}}
\newcommand\p{{\scriptscriptstyle\rm P}}
\newcommand\rs[1]{{\scriptscriptstyle\rm #1}}
\newcommand\pp{{\vphantom\p}}
\newcommand\pdag{{\vphantom \dagger}}
\newcommand\normal[1]{\,{:} #1 {:}\,}

\layoutstyle{8x11double}

\begin{document}

\title{Single-Particle Excitations Generated by Voltage Pulses}



\author{Fabian Hassler}{
  address={Institute for Theoretical Physics, ETH Zurich, 8093 Zurich,
Switzerland}
}

\author{Bruno K\"ung}{
  address={Solid State Physics Laboratory, ETH Zurich, 8093 Zurich,
Switzerland}
}

\author{Gordey B. Lesovik}{
  address={L.D.\ Landau Institute for Theoretical Physics RAS,
117940 Moscow, Russia}
}

\author{Gianni Blatter}{
  address={Institute for Theoretical Physics, ETH Zurich, 8093 Zurich,
Switzerland}
}

\begin{abstract}
  We analyze properties of excitations due to voltage pulses applied to a 1D
  noninteracting electron gas, assuming that the integral of the voltage over
  time is equal to the unit of flux. We show that the average charge transfer
  due to such pulses does not depend on the pulse shape. For pulses with a
  Lorentzian profile, we prove the single-particle nature of the electron
  and the hole excitations.
\end{abstract}

\maketitle

%%%%%%%%%%%%%%%%%%%%%%%%%%%%%%%%%%%%%%%%%%%%
%% MAINMATTER
%%%%%%%%%%%%%%%%%%%%%%%%%%%%%%%%%%%%%%%%%%%%

\section{Introduction}

Recently, much interest has concentrated on single-particle sources
feeding devices with individual electrons \cite{lee:95,levitov:96,
ivanov:97,lebedev:05,keeling:06,feve:07,keeling:08,mahe:08}. The action
of a voltage pulse can be seen as a unitary transformation $\hat{U}$ (a
scattering operator) acting on the degenerate Fermi sea $|\Phi_\F \rangle$
of the one-dimensional system.  In this work, we will show that for properly
designed pulses this state is free of entangled particle-hole pairs but
instead involves a pure one-particle excitation on top of a complete
Fermi sea. This result was first derived by Keeling \emph{et al.} who
describe the action of the voltage pulse on the single particles states
and argue that many-particle excitations are absent \cite{keeling:06}.
Here, we present an alternative proof which describes the application
of the voltage pulse as a (second-quantized) scattering matrix $\hat U$
such that the nature of the excitation can be shown rigorously within the
many-body setup.  First experiments involving single-electron sources were
performed recently by F\`eve \emph{et al.} \cite{feve:07} (for a theoretical
description see Ref.~\cite{keeling:08}), using a quantum dot in a
quantum Hall sample at integer filling.

\begin{figure}
  \centering
  \includegraphics{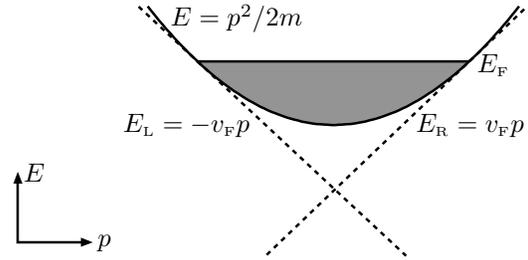}
  \caption{For low temperatures and excitation energies, the quadratic
  dispersion relation $E=p^2/2m$ can be expanded around the Fermi
  surface. In one dimension, this leads to two independent branches
  $E_\rs{R/L} = \pm v_\F p$ with $v_\F = p_\F/m$ (up to an irrelevant energy
  shift). Here, the subscript R (L) refers to a particle moving to the right
  (left).}\label{fig:lin_spec}
\end{figure}

The calculations in this chapter are performed using a linear dispersion
relation $E_\rs{R/L} = \pm v_\F p$ valid close to the Fermi points $\pm p_\F$
[cf.\ Fig.~\ref{fig:lin_spec}], where $v_\F$ ($p_\F$) denotes the Fermi
velocity (momentum). In one dimension, the linear-spectrum approximation
disconnects the dispersion relation of the electrons into right (R) and
left (L) moving branches. Here, we are interested in generating, via a
voltage pulse, a particle in the right-moving branch together with a hole
in the left-moving branch which leads to a net current in the device. The
voltage pulse (together with the linear-spectrum approximation) conserves
the number of particles in each branch individually. In order to be able
to describe the above process, we need to introduce an infinitely deep
Fermi sea (by setting the lower bound of the momentum in each branch to
$-\infty$) \cite{delft:98}. We then proceed as follows: First, we prove
that the average charge transmitted through the wire is related to the
voltage integrated over time, i.e., each flux-quantum produces one electron
on average (Secs.~\ref{sec:volt} and \ref{sec:total}). Subsequently,
we concentrate on the right-moving branch and show how a voltage pulse
can be described via a (unitary) scattering operator $\hat{U}$ and that
for a unit-flux Lorentzian pulse the excitation is of a single-particle
nature (Sec.~\ref{sec:1particle}).  Finally, to complete the picture,
we turn our attention to the left-moving branch and demonstrate that the
single-particle excitation in the right-moving branch is accompanied by
a single hole in the left-moving branch (Sec.~\ref{sec:1hole}).

\section{Voltage pulses}\label{sec:volt}

\begin{figure}
  \centering
  \includegraphics{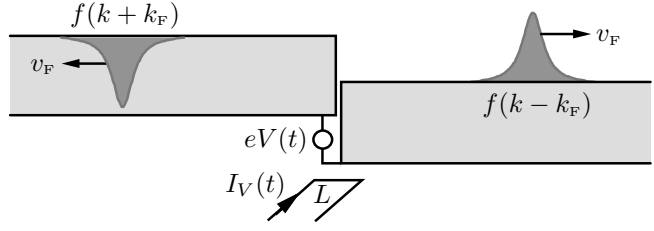}
  \caption{Due to the application of a time-dependent, external current
  $I_V(t)$, a time-dependent voltage $V(t)$ acts on the electrons in
  the quantum wire. Assuming that the voltage pulse has unit-flux and a
  Lorentzian shape, the shakeup of the Fermi sea creates a single-particle
  excitation with wave function $f(k-k_\F)$ which moves to the
  right and simultaneously a hole excitation of the same type which
  moves to the left. Note that the electron-hole pair is entangled
  \cite{beenakker:03}.}\label{fig4:setup}
\end{figure}

Consider a quantum wire where voltage pulses $V(t)$ can be applied over a
small region around $x=x_V$. Such voltage pulses act on the right-moving
electrons with dispersion $E_\rsR = v_\F p$ and produce electronic
excitations in the quantum wire [cf.\ Fig.~\ref{fig4:setup}].%
\footnote{Note that in the experiment \cite{kozhevnikov:00}
on photon-assisted noise \cite{lesovik:94} instead of a voltage pulse
localized at a specific position a bias voltage over the whole sample was
used. This procedure might not work in our case as our effect depends on
the details of the voltage drop across the sample.}
If the time dependence of the voltage is slow compared to
the transition time of an electron through the region with the voltage drop,
the potential can be considered as quasistatic.  In this case, the effect of
the voltage pulse can be incorporated in a phase factor $\exp [i \phi(t -
x/v_{\rm\scriptscriptstyle F}) \Theta(x-x_V)]$ multiplying the wave function,
where the phase
\begin{equation}\label{eq4:phase} 
  \phi(t) = \frac{e\Phi(t)}{\hbar c}
  = \frac{e}{\hbar} \int_{-\infty}^t \!dt' \, V(t')
\end{equation}
is proportional to the flux $\Phi(t)$, $\Theta(x)$ denotes the unit-step
function, and $e>0$ is the unit of charge. The right-moving scattering
state $\psi_{\rsR,k} (x;t) = \exp [i k (x- v_{\rm\scriptscriptstyle F} t)
+ i \phi(t- x/ v_{\rm\scriptscriptstyle F}) \Theta(x-x_V) ]$ is a solution
of the linearized time-dependent Schr\"odinger equation
\begin{equation}\label{eq4:time_schroedinger}
  \Bigl[ i \hbar (\partial_t + v_\F \partial_x) +
  \hbar v_\F \phi(t) \delta(x-x_V) \Bigr] \psi_{\rsR,k}(x;t) =0
\end{equation}
including the time-dependent voltage $V(t)$ via $\phi(t)$
\cite{lesovik:94,keeling:06}. Left of the position of the voltage pulse,
$x<x_V$, $\psi_{\rsR,k}$ is a plane wave with well-defined energy $\hbar
v_{\rm\scriptscriptstyle F} k$. On the right of the position of the voltage
pulse, $x>x_V$, the wave function $\psi_{\rsR,k}(x;t)$ can be decomposed
into energy eigenmodes $\exp[ik(x-v_{\rm\scriptscriptstyle F} t) ]$ of
the free Hamiltonian
\begin{equation}\label{eq4:fourier}
  \psi_{\rsR,k} ( x >x_V;t) = 
  \int \frac{dk'}{2\pi} 
  U(k'-k) e^{i k' (x - v_{\rm\scriptscriptstyle F} t)}
\end{equation}
where the transformation kernel 
\begin{equation}\label{eq4:u} U(q) = v_{\rm\scriptscriptstyle F} \int dt 
  \,e^{i \phi(t)
  + i q v_{\rm\scriptscriptstyle F} t}
\end{equation} 
is the Fourier transform of the phase factor $\exp [i \phi(t)]$.  Here, we
are interested in applying integer flux pulses such that $\phi(t \rightarrow
\infty) \in 2\pi \mathbb{N}$, as noninteger flux pulses do not produce
clean single-particle excitations and lead to logarithmic divergences in
the noise of the transmitted charge \cite{lee:93,levitov:96}. In the case
of integer flux pulses, the long time asymptotics $\exp [i \phi(t \to \pm
\infty)]\to 1$ generates a Dirac delta function $2\pi \delta (q)$
in $U(q)$.  The remaining part
\begin{equation}\label{eq4:ureg}
  U^\text{reg} (q) = U(q) - 2\pi \delta(q) =
  v_{\rm\scriptscriptstyle F} \int dt  \bigl[ e^{i \phi(t)} -1 \bigr]
  e^{i q v_{\rm\scriptscriptstyle F} t}
\end{equation}
is finite for a localized (in time) voltage pulse.  The transformation
$U(k'-k)$ describes the scattering amplitude for the transition from a
momentum state $k$ (for $x<x_V$) to the state $k'$ (for $x>x_V$) due to the
application of the voltage pulse. The statement that the wave function is
in the momentum state $k'$ holds in the asymptotic region $x-x_V\gg v_\F
\tau$ with $\tau$ the typical timescale associated with the dynamics of
the voltage pulse.

As we are interested in the region with $x>x_V$, the scattering states
originating from the right do not enter the region where the voltage pulse
is applied and therefore are unperturbed
\begin{equation}\label{eq4:right}
 \psi_{\rsL,k} (x> x_V; t) = e^{- i k (x+ v_{\rm\scriptscriptstyle F} t)}
\end{equation}
without any shift in energy.%
\footnote{In the linear-spectrum approximation particles are not
back-reflected due to the voltage-pulse. The hole appears for $x<x_V$,
see later Sec.~\ref{sec:1hole}.}
The time-dependent field operator
in the region $x>x_V$ behind the voltage pulse is given by
\[
 \Psi (x>x_V; t) = \int \frac{dk}{2\pi} \bigl[ 
 \psi_{\rsR,k} (x;t) a_{\rsR,k} + 
 \psi_{\rsL,k} (x;t) a_{\rsL,k} \bigr] ;
\]
here, $a_{\rsR,k}$ ($a_{\rsL,k}$) denote fermionic annihilation operators
for states moving to the right (left), i.e., incoming from the left
(right) reservoir.  Averaging the current operator
\[
I (x >
x_V;t) = \frac{ie \hbar}{2m}
\bigl[\Psi^\dagger(x;t) \partial_x \Psi(x;t) - \text{H.c.} \bigr]
\]
over the Fermi reservoirs assuming the same Fermi distribution $n(k)
= n_\rsR(k) = n_\rsL (-k)$ initially at the far left and right of the
interaction region and integrating the ensemble averaged current over time,
we obtain the transmitted charge
\begin{equation}\label{eq4:mean_charge}
 \langle -Q/e \rangle = 
  \int \frac{dk' \, dk}{(2\pi)^2} \, K(k'-k) n(k)
\end{equation}
with the kernel
\begin{equation}\label{eq4:kq_def_nomol}
K(q) = |U^\text{reg} (q)|^2 +
4\pi \delta(q) \text{Re} \bigl[ U^\text{reg} (0) \bigr];
\end{equation}
a more careful calculation shows that $\text{Re} [U^\text{reg} (0)]$ has
to be replaced by the symmetric limit $\text{Re} [U^\text{reg} (0^+)
+ U^\text{reg} (0^-)]/2$ in those cases where $U^\text{reg}(q)$
has a discontinuous real part near $q=0$.%
\footnote{This is the case for the unit-flux Lorentzian voltage pulses
which produce the single-particle excitations.}
Equivalently, the kernel $K(q)$ can be defined as
\begin{equation}\label{eq4:k}
  K(q) = v_{\rm\scriptscriptstyle F}^2 \int dt dt' 
  \bigl[ e^{i \phi(t) - i\phi(t')} -1
  \bigr] e^{i q v_{\rm\scriptscriptstyle F} (t-t')}.
\end{equation}
Note that $\int d q \,K(q) =0$, which can be obtained from the fact that
integrating over $q$ in Eq.~(\ref{eq4:k}) leads to a $\delta$-function
imposing $t=t'$, such that the term in the rectangular bracket vanishes.

\section{Total transferred charge}\label{sec:total}

In this section, we will show that charge transport caused by an integer-flux
pulse at zero temperature depends only on the number of flux units it
contains, independent of the pulse shape.  In the last section, we have
derived the formula [cf.\ Eq.~(\ref{eq4:mean_charge})]
\begin{equation}
  \label{eq4:qe_secondquantized_nomol}
  \langle -Q/e\rangle = \int \frac{d k' \, dk}{(2\pi)^2} \, K(k'-k) n(k)
\end{equation}
providing the charge $\langle - Q/e \rangle$ transferred in a process
involving an integer-flux voltage-pulse with arbitrary shape.%
\footnote{Note that the integrals in (\ref{eq4:mean_charge}) cannot be
performed in arbitrary order. Calculating first the integral over $k'$,
a vanishing transferred charge is obtained. Similarly, the charge vanishes
in the case of a finite Fermi sea. Only in the case of an infinitely deep
Fermi sea a finite answer is obtained. This reflects the fact that the
number of particles in right-moving states is conserved in a system with
a finite Fermi sea. In reality, the quadratic spectrum connects the left-
with the right-going branch and effectively promotes one left-going electron
to a right-going state.}
To proceed, we assume the thermal energy $k_\textrm{B} \vartheta$ to be
smaller than any other energy scale, so we can employ the zero temperature
distribution $n(k)= \Theta(k_\F-k)$. Inserting the distribution function
into (\ref{eq4:qe_secondquantized_nomol}) and changing the integration
variable $k$ to $q=k'-k$ yields
\begin{equation}
  \label{eq4:qe2nd_variablechange} \langle -Q/e \rangle =\int \! \frac{d
  k'}{2\pi} \int \! \frac{d
    q}{2\pi} K(q) \Theta(k_\F-k'+q).\nonumber
\end{equation}
We want to obtain a form for the transmitted charge, which only depends
on the first term of Eq.~(\ref{eq4:kq_def_nomol}) for $K(q)$. We do that
by analyzing the $q$ integration at fixed values of $k'$. If $k' > k_\F$,
the $\delta$-function in $K(q)$ lies outside of the integration range and
the second term in Eq.~(\ref{eq4:kq_def_nomol}) does not contribute. In
the other case $k'<k_\F$, we first use the identity $\int d q \, K(q)
\Theta(k_\F-k'+q) = - \int d q \, K(q) \Theta(k'-k_\F-q)$ which follows
from $\int dq K(q) = 0$, and again the $\delta$-function lies outside the
integration range. In summary, we obtain
\begin{align*}
    \langle -Q/e \rangle = \int \! \frac{d k' \, dq}{(2\pi)^2} 
    |  U^\text{reg}(q&)|^2 \Big[\Theta(k'-k_\F) \Theta(k_\F-k'+q) \\
    &-\Theta(k_\F-k') \Theta(k'-k_\F-q) \Big]. 
\end{align*}
The integration of the rectangular bracket over $k'$ yields a factor $q$
and we arrive at the formula
\begin{equation}
  \label{eq4:qe2nd_alternative}
  \langle -Q/e \rangle = 
  \int \! \frac{d q}{(2\pi)^2} \, q \, |U^\text{reg}(q)|^2
\end{equation}
for the average transmitted charge. Here, the factor $q$ is
proportional to the number of particles which can be excited
above the Fermi edge by providing them the momentum $q$ and
$|U^\text{reg}(q)|^2$ denotes the probability for such an event to happen.
Equation~(\ref{eq4:qe2nd_alternative}) has a number of advantages over the
expression in Eq.~(\ref{eq4:qe_secondquantized_nomol}). The integration
over the Fermi sea is already performed and only one integration over
the momentum $q$ of the excitations remains. Furthermore, it involves the
kernel $U^\text{reg}(q)$ [instead of $K(q)$] which is directly related to
the Fourier transform of the phase acquired by the voltage pulse $V(t)$
[see Eq.~(\ref{eq4:ureg})].  Inserting the definition (\ref{eq4:ureg}) into
Eq.~(\ref{eq4:qe2nd_alternative}) and performing the momentum integration
yields the Fourier-transformed expression
\begin{align*}
\langle -Q/e\rangle &= \int \frac{d t}{2\pi} 
 \bigl[e^{-i\phi(t)}-1\bigr] \frac{d \,}{i d t} \bigl[e^{i\phi(t)}-1\bigr] \\
 &= \int \frac{dt}{2\pi} 
 \, \dot{\phi^{\vphantom{\Phi}}}(t) \bigl[e^{-i\phi(t)}-1\bigr] e^{i\phi(t)},
\end{align*}
where the boundary terms vanish because the voltage pulse carries an
integer flux with $\exp[i\phi(\pm \infty)] =1$. Changing the integration
variable from $t$ to $\phi$ and performing the integration yields
\begin{equation}
  \label{eq4:qe2nd_steptwo}
  \langle-Q/e\rangle= -\frac{\phi(+\infty)-\phi(-\infty)}{2\pi}
\end{equation}
In this form, it is visible that the average transmitted charge $\langle
-Q/e \rangle $ does not depend on the pulse shape but only on the winding
number of $e^{i\phi(t)}$. For example, a negative unit-flux pulse produces
a phase which is decreasing in time with $\phi(+\infty)-\phi(-\infty) =
-2\pi$. Therefore, on average one electron with charge $-e$ transferred.

\section{Single-particle excitation}\label{sec:1particle}

In a time-dependent basis of right-moving position eigenstates $\langle
x;t |x_\rsR \rangle = \delta (x-v_\F t)$, labeled by the retarded
position $x_\rsR = x-v_\F t$, the effect of the voltage pulse is easy to
incorporate. In a first step, we will concentrate on these right-moving
states (in this section, we drop the subscript R).  The states pick up
a scattering phase $\exp[i\phi(-x_\rsR/v_\F)]$ when they pass through
$x=x_V$, with $\phi(t) = (e/\hbar) \int_{-\infty}^t \!\! V(t')d t'$.
The operator $\hat{U}$, which transforms the initial states $| x_\rsR
\rangle$ at $x<x_V$ into the corresponding voltage-driven states at $x>x_V$,
can then be written as the unitary operator
\begin{equation}
  \label{eq4:unitary_transformation}
  \hat{U} = \exp \left[i \int \! \!d
  x_\rsR \,\phi(-x_\rsR/v_\F)\Psi^\dagger(x_\rsR) \Psi(x_\rsR) \right],
\end{equation}
where $\Psi(x_\rsR)$ is a second-quantized fermion operator which
annihilates a particle in the state $| x_\rsR \rangle$. Applying a
unit-flux Lorentzian voltage pulse
\begin{equation}\label{eq4:lorentz}
  V(t)= -\frac{\hbar}{e} \frac{2 v_\F \xi}{(v_\F t)^2+\xi^2}
\end{equation}
with width $\xi$, adds a phase
\begin{equation}\label{eq4:phase_lorentz}
  \phi(t) = -[\pi + 2 \arctan(v_\F t/\xi)]
\end{equation}
to the right going scattering states and thereby excites the Fermi sea. Here,
we want to prove that
\begin{equation}
  \label{eq4:UPhi}
  \hat U | \Phi_\F \rangle = A^\dagger | \Phi_\F \rangle
\end{equation}
where the single-particle creation operator $A^\dagger$ is defined via
\begin{equation}
  \label{eq4:A_def} A^\dagger = \int \! \frac{d k}{2\pi} f(k-k_\F) 
  a^\dagger_k 
\end{equation}
and the creation operators $a_k^\dagger$ are associated with right-moving
plane waves $\exp(ikx_\rsR)$; the amplitude
\begin{equation} 
  \label{eq4:f}
  f (\varkappa) =
  \sqrt{4\pi\xi} e^{-\xi \varkappa} \Theta(\varkappa)
\end{equation}
is (up to normalization $N$) given by $f(\varkappa) = N \, U^\text{reg}
(\varkappa)$ [cf.\ Eq.~(\ref{eq4:ulorentz})].  Equation (\ref{eq4:UPhi})
tells us that a unit-flux Lorentzian pulse creates a clean single-particle
excitation, i.e., a product state of the filled Fermi sea $|\Phi_\F\rangle$
and a particle with wave function $f(k-k_\F)$ propagating independently of
$|\Phi_\F\rangle$. This excitation can then be described as a Lorentzian
wave packet in the first-quantized language.

We prove Eq.~(\ref{eq4:UPhi}) by calculating general correlators of the form
\begin{equation}
  \label{eq4:correlator}
  \langle \Phi|
  \normal{a_{k_{1}}^\dagger a_{k_{1}}^{\vphantom{\dagger}} 
  \cdots a_{k_{n}}^\dagger
  a_{k_{n}}^{\vphantom{\dagger}}} |\Phi\rangle
\end{equation}
and show that they are equal for the states $|\Phi_A \rangle = A^\dagger
|\Phi_\F\rangle$ and $|\Phi_U\rangle =\hat U |\Phi_\F\rangle$.  Here and
below $\normal{\mathcal{O}}$ denotes the normal ordering of the operator
$\mathcal{O}$ defined via $\normal{\mathcal{O}}=\mathcal{O}-\langle \Phi_\F
| \mathcal{O} | \Phi_\F \rangle$ which has to be applied to render the
correlators finite.%
\footnote{For fermionic operators $a_k$, the normal ordering
$\normal{a_k^\dagger a_k^\pdag}$ is given by $a_k^\dagger a_k^\pdag$
for $k>k_\F$ and $- a_k^\pdag a_k^\dagger$ for $k<k_\F$.}
Since the plane waves associated with the creation operators $a_k^\dagger$
form a complete basis of the one-particle Hilbert space of right-moving
electrons, an arbitrary many-particle state $|\Phi\rangle$ (assuming a
fixed number of particles) can be fully specified by the set of correlators
(\ref{eq4:correlator}). While the calculation of the correlators for the
state $|\Phi_A\rangle = A^\dagger |\Phi_\F\rangle$ is simple, it is not at
all straightforward for $|\Phi_U\rangle$. The key part in our derivation
is to show that the correlators are zero for $n\geq 2$ which proves the
single-particle nature of the excitation.

Using the explicit form (\ref{eq4:unitary_transformation}) of the unitary
scattering operator, the transformation
\begin{equation}
  \label{eq4:a_transform}
  \begin{split}
  \hat U^\dagger a_k \hat U &= 
  a_k + \int \! \frac{d k'}{2\pi} U^\text{reg} (k - k') a_{k'} \\
  &= \int \! \frac{dk'}{2\pi} U(k-k') a_{k'}
\end{split}
\end{equation}
of the annihilation operator $a_k$ can be derived straightforwardly,
where the kernel
\begin{equation}\label{eq4:ulorentz}
  U^\text{reg}(q) =
  -2\pi (2\xi) e^{-\xi q} \Theta(q) 
\end{equation}
has been introduced before [cf.\ Eq.~(\ref{eq4:ureg})].  Using this relation,
we calculate the one-particle matrix element%
\footnote{We have evaluated the creation and annihilation operator at
different momenta in order to avoid a subtraction of two divergences.}
\begin{multline}\label{eq4:PhiUaaUPhi_null}
  \langle \Phi_U | \normal{a^\dagger_{k'}  
  a^{\vphantom{\dagger}}_k } |\Phi_U \rangle = \int_{-\infty}^{k_\F} 
  \! \frac{d
  k''} {2 \pi}[U^*(k'-k'') U(k-k'')\\
    - (2\pi)^2\delta(k'-k'')
    \delta(k-k'')].
\end{multline}
Inserting the particular form (\ref{eq4:ulorentz}) of $U(q)$ for a Lorentzian
pulse into Eq.~(\ref{eq4:PhiUaaUPhi_null}), the matrix element yields
\begin{equation}
  \label{eq4:PhiUaaUPhi}
   \langle \Phi_U | \normal{a^\dagger_{k'}  
   a^{\vphantom{\dagger}}_k} |\Phi_U\rangle = 
   f^*(k'-k_\F)
  f(k-k_\F)
\end{equation}
with $f(k-k_\F)$ defined in Eq.~(\ref{eq4:f}).  The one-particle
correlator (\ref{eq4:PhiUaaUPhi}) is the same as the one generated by
the state $|\Phi_A\rangle$. 

Having shown the equivalence of the states $|\Phi_A\rangle$ and
$|\Phi_U\rangle$ on the one-particle level, we want to proceed showing
that this relation also holds for higher-order correlators. In a first
step, we show that (\ref{eq4:correlator}) vanishes, whenever one of the
$k$'s is smaller than the Fermi wave-vector $k_\F$, thereby we use the
property that $U(q) \propto \Theta(q)$ (I). Physically, this means that
there is no hole excitation in the system. Then, we show that the correlator
(\ref{eq4:correlator}) also vanishes whenever there are two entries $k$ which
are larger than the Fermi wave-vector. To this end, we use the fact that
$U(q) \propto e^{-q}$ (II). Properties (I) and (II) uniquely restrict the
corresponding voltage pulse to the unit-flux Lorentzian form.  Thus, this is
the only pulse shape which may generate a single-particle excitation.%
\footnote{This conclusion relies on the fact that $U(q)$ only depends
on the transferred momentum $q$, which is valid in the present setup of
a voltage driving the single-particle excitation. Allowing $U(k_1,k_2)$
to depend both on the incoming $k_1$ and on the outgoing $k_2$ momentum,
other shapes become possible, see for example Ref.~\cite{keeling:08}.}

First, we show that the correlator (\ref{eq4:correlator}) vanishes
if one $k_j<k_\F$. Due to the normal ordering, the creation operator
$a^\dagger_{k_j}$ can be moved to the very right where it acts on
$|\Phi_U\rangle$ and vanishes,
\begin{equation}
  \label{eq4:show}
  \begin{split}
  a_{k_j}^\dagger |\Phi_U\rangle &= \hat U \hat U^\dagger
  a_{k_j}^\dagger \hat U |\Phi_\F\rangle \\
  &= \hat U \int \! 
  \frac{d k'} {2\pi} U^*(k_j -k') a^\dagger_{k'} |\Phi_\F\rangle = 0;
\end{split}
\end{equation}
here, we have used the adjoint of Eq.~(\ref{eq4:a_transform}). The last
equality follows from the fact that $k'>k_\F$ and $k_j<k_\F$, together
with the property that the kernel $U(q)$ only increases the energy [as it
is proportional to $\Theta(q)$].

Having found that there are no hole excitations present, we have to check
the single-particle nature of the state $|\Phi_U\rangle$. For this, we
assume that there are two $k$'s, $k_j$ and $k_l$, which are larger than
the Fermi wave-vector $k_\F$. Here, we show that $a_{k_j} a_{k_l} \hat U
|\Phi_\F\rangle=0$ which makes all correlators with more than one particle
above the Fermi sea vanish, effectively proving that the excitation above
the Fermi sea is of single-particle nature.  Applying the annihilation
operators on the excited state $|\Phi_U\rangle$ yields
\begin{align*}
  a_{k_j} a_{k_l} |\Phi_U\rangle =   
  \hat U &\hat U^\dagger a_{k_j} \hat U \hat U^\dagger  a_{k_l} \hat U
  |\Phi_\F\rangle \\
 =\hat U &\biggl(
 a_{k_j} + \! \int \frac{d k}{2\pi} U^\text{reg} (k_j - k)
    a_{k} \biggr) \\
  \times&\biggl( a_{k_l} + \! \int \frac{d k}{2\pi} U^\text{reg} (k_l - k)
    a_{k} \biggr) |\Phi_\F\rangle. 
\end{align*}
Next, we use the fact that the contribution from terms containing an
annihilation operator $a_k$ with $k>k_\F$ vanishes. This includes both the
isolated $a_{k_j}$ and $a_{k_l}$, as well as the parts of the integrals with
$k > k_\F$. Moreover, we insert the explicit form Eq.~(\ref{eq4:ulorentz})
of $U^\text{reg}(q)$ and obtain
\begin{align}
  \label{eq4:aaUPhi}
  a_{k_j} a_{k_l} \hat U |\Phi_\F \rangle = \! \hat U 
  &\biggl(2\xi e^{-\xi(k_j-k_\F)} \int_{-\infty}^{k_\F} \!\!
    d k \,    e^{-\xi(k_\F - k)} a_{k} \biggr) \nonumber\\
  \times &\biggl(2\xi e^{-\xi(k_l-k_\F)}
    \int_{-\infty}^{k_\F} \!\! d k \,
    e^{-\xi(k_\F - k)} a_{k} \biggr)  |\Phi_\F \rangle \nonumber\\
    = 0. & 
\end{align}
The last equality follows from the fact that both factors are proportional to
the same fermionic operator $\int_{-\infty}^{k_\F} d k \, \exp[-\xi(k_\F-k)]
a_k$.  Due to Fermi statistics, the corresponding state can be occupied
by only one particle.%
\footnote{Keeling \emph{et al.} denote this decisive property as $\hat U$
being of rank~1.  }

Summing up, we have shown that the correlator (\ref{eq4:correlator}) is
given by (\ref{eq4:PhiUaaUPhi}) for $n=1$. For $n\geq2$, the correlator
vanishes as there are either more than two $k$'s larger than the Fermi
momentum [using Eq.~(\ref{eq4:aaUPhi})] or at least one $k$ is below $k_\F$
[using Eq.~(\ref{eq4:show})]. We obtain
\begin{align*}
  \langle
  \Phi_U | \normal{  a_{k_{1}}^\dagger a^\pdag_{k_1} 
  \cdots a_{k_{n}}^\dagger a_{k_n}^\pdag} &
  |\Phi_U \rangle = |f(k_1-k_\F)|^2 \delta_{n,1} \\&=
  \langle\Phi_A| 
  \normal{  a_{k_{1}}^\dagger a^\pdag_{k_1} 
    \cdots a_{k_{n}}^\dagger a_{k_n}^\pdag}
  | \Phi_A \rangle, 
\end{align*}
thus completing the proof of Eq.~(\ref{eq4:UPhi}).

\section{Hole excitation}\label{sec:1hole}

The result (\ref{eq4:UPhi}) is yet incomplete, since it is not clear where
the additional particle $A^\dagger$ comes from; the complete result must
satisfy particle number conservation. To understand how this problem
is solved, we have to recall that we described the electron system as
\emph{two independent} systems of basis states, right- and left-moving,
respectively. This reflects the physical picture that there are two
mutually-independent sets of states near the two Fermi points of the
one-dimensional electron gas. Above, the action of the Lorentzian pulse on
the set of \emph{right-moving} states was determined.%
\footnote{The fact that an operator $\hat U$ which evidently commutes
with the particle number operator generates an additional particle is
not a contradiction since the Fermi sea $|\Phi_\F\rangle$ contains an
infinite number of particles.  Similar ideas are also used in the context
of bosonization of a 1D fermionic system, see Ref.~\cite{delft:98}.}
The empty state is instead found in the set of \emph{left-moving} states,
i.e., the voltage pulse creates a hole near the point $-k_\F$, propagating
in the direction opposite to that of the particle-like excitation. In
the following, we will demonstrate how the problem of describing the hole
excitation in the left-moving branch can be mapped back on the problem of
generating an electron in the right-moving branch in two steps. First,
a parity transformation is applied which changes the branch but also
reverses the sign of the voltage pulse. To return back to the initial
sign of the voltage pulse, we additionally employ a particle-hole
transformation which effectively inverts the unit of charge.%
\footnote{The 1D Fermi system with linearized momentum is particle-hole
symmetric with the symmetry operation $\Psi^\dagger (x_\rsR) \to e^{-i
p_0 x_\rsR}\, \Psi( x_\rsR)$ in the branch of right-moving states. Note
that the symmetry operation inverts the momentum $k \to p_0- k$; here we
choose $p=2k_\F$.}
Together, these transformations map the problem of the generation of the
hole on the problem of the generation of the electron. In the following,
we describe these transformations more formally.

Below, we want to prove that the voltage pulse $\hat U_\rsL$ acting on the
ground state $|\Phi_\F\rangle_\rsL$ generates a single-hole, i.e., defining
the operator $B$ via 
\begin{equation}\label{eq4:hole}
  \hat U_\rsL |\Phi_\F \rangle_\rsL = B | \Phi_\F \rangle_\rsL,
\end{equation}
this operator generates a single-hole
\begin{equation}\label{eq4:b}
  B = \int \! \frac{dk}{2\pi}\, f (k+k_\F) a_{\rsL,k}
\end{equation}
in a specific state with amplitude $f(k+k_\F)$. The scattering matrix for
the left-moving states is given by
\begin{equation}
  \label{eq4:unitary_transL}
  \hat{U}_\rsL = \exp \left[-i \int \! \!d
  x_\rsR \,\phi(x_\rsL/v_\F)
  \Psi_\rsL^\dagger(x_\rsL) \Psi_\rsL(x_\rsL) \right],
\end{equation}
with $x_\rsL = x+v_\F t$ and the different sign of the phase with
respect to Eq.~(\ref{eq4:unitary_transformation}) appears due to the
fact that the left-moving particles traverse the voltage in the opposite
direction.  In a first step, we observe that instead of considering the
left-moving states, we can equivalently perform a parity transformation
$x \to x^\p =-x$ and then deal with the right-moving states. The parity
transformation changes the branch $E_\rsL \to E_\rsL^\p= E_\rsR^\pp$
as well as the momentum $k \to k^\p = -k$, such that the annihilation
operators transform according to $a^\pp_{\rsL,k} \to a_{\rsL,k}^\p =
a^\pp_{\rsR,-k}$ and $\hat U^\pp_\rsL \to \hat U^\text{P}_\rsL = \hat
U^{\pp\dagger}_\rsR$ [cf.\ Eqs.~(\ref{eq4:unitary_transformation})
and (\ref{eq4:unitary_transL})]. Applying the parity transformation,
Eq.~(\ref{eq4:hole}) reads
\begin{equation}\label{eq4:holeP}
  \hat U^\dag_\rsR |\Phi_\F\rangle_\rsR = B^\p |\Phi_\F \rangle_\rsR.
\end{equation}

Next, we implement a particle-hole transformation $a^\dag_{\rsR,k_\F+\varkappa}
\to \bar a^\dag_{\rsR,k_\F+\varkappa} = a^\pdag_{\rsR,k_\F-\varkappa}$ on the
branch of the right-moving states which has been chosen in such a way as to
keep the excitation energy $v_\F \varkappa$ (and thereby the Fermi level $k_\F$)
as well as the anti-commutation relations unchanged. The particle-hole
transformation can be implemented on the level of field operators
\begin{align*}
  \Psi_\rsR^\dagger(x_\rsR) \to \bar \Psi_\rsR^\dagger(x_\rsR) &=
  \int \! \frac{d \varkappa}{2\pi} e^{-i (k_\F + \varkappa) x_\rsR} \, \bar
  a^\dagger_{\rsR, k_\F +\varkappa} \\
  &= e^{- 2 i k_\F x_\rsR} \int \! 
  \frac{d\varkappa}{2\pi} e^{i (k_\F - \varkappa) x_\rsR} \, 
    a^\pdag_{\rsR, k_\F -\varkappa} \\
  &= e^{- 2 i k_\F x_\rsR} \, \Psi_\rsR (x_\rsR).
\end{align*}
Using this relation, we can transform the scattering operator
\begin{align}\label{eq4:ubar}
  \overline{\hat U_\rsR^\dag}&= 
  \exp [-i\! \int \!d x_\rsR \,\phi(-x_\rsR/v_\F)
  \bar \Psi_\rsR^\dagger(x_\rsR)
  \bar \Psi_\rsR^\pdag (x_\rsR) ] \nonumber\\
  &=  \exp [-i\! \int \!d x_\rsR \,\phi(-x_\rsR/v_\F) 
  \Psi_\rsR^\pdag(x_\rsR)
  \Psi_\rsR^\dagger(x_\rsR) ] \nonumber\\
  & = c \, \exp [i\! \int \!d x_\rsR \,\phi(-x_\rsR/v_\F) 
  \Psi_\rsR^\dagger(x_\rsR)
    \Psi_\rsR^\pdag(x_\rsR) ] \nonumber\\
  & = c \, \hat U_\rsR,
\end{align}
where $c$ is a phase factor which we will set equal to one in
the following.%
\footnote{Note that the phase factor $c$ appearing when commuting the
operators in Eq.~(\ref{eq4:ubar}) is undefined as a factor $\delta(0)$
appears. This fact poses no further problems as an overall phase factor
in a state is not measurable.}
After the particle-hole transformation, Eq.~(\ref{eq4:holeP}) reads
\begin{equation}\label{eq4:holePH}
  \hat U_\rsR |\Phi_\F\rangle_\rsR = \overline{B^\p} |\Phi_\F\rangle_\rsR.
\end{equation}
The comparison of Eq.~(\ref{eq4:holePH}) with Eq.~(\ref{eq4:UPhi}) yields
$\overline{B^\p} = A^\dagger$ from where $B$ can be obtained by inverting
and subsequent usage of Eq.~(\ref{eq4:A_def}). Finally, we obtain
\begin{align}\label{eq4:bequala}
  B= \overline{A^\dagger}^\p
  &=  \Biggl[ \int \! \frac{d k}{2\pi} f(k-k_\F) 
  \bar {a}_{\rsR, k }^\dagger 
  \Biggr]^\p \nonumber\\
   &= \int \! \frac{d \varkappa}{2\pi} f(\varkappa) 
   a^{\pdag\p}_{\rsR,k_\F -\varkappa}
    = \int \! \frac{d \varkappa}{2\pi} f(\varkappa) 
          a^\pdag_{\rsL, \varkappa-k_\F} \nonumber\\
   & = \int \! \frac{d k}{2\pi} f(k+k_\F) 
             a^\pdag_{\rsL, k}
\end{align}
proving (\ref{eq4:hole}). The application of the voltage pulse creates a
single-hole excitation with wave function $f(k+k_\F)$ which moves to the
left. Similar to the case of the particle excitation in the right-moving
states (\ref{eq4:UPhi}), this state can be described in first-quantized
language.

\section{Conclusion}

We have shown how the application of a unit-flux Lorentzian voltage pulse
leads to single-particle excitations, an electron moving to the right
and a hole moving to the left. These excitations can be described in a
first-quantized language. For general integer-flux pulses, it was proven
that the average transported charge is determined by the integral of the
voltage alone, independent of the pulse shape. The proof has been outlined
within a linear-spectrum approximation and using an infinitely-deep Fermi
sea. The effect of the quadratic spectrum on the present results remains
an interesting problem for further studies.

%%%%%%%%%%%%%%%%%%%%%%%%%%%%%%%%%%%%%%%%%%%%%%%%
%% BACKMATTER
%%%%%%%%%%%%%%%%%%%%%%%%%%%%%%%%%%%%%%%%%%%%%%%%

\begin{theacknowledgments}
  We acknowledge financial support from the CTS-ETHZ, the Swiss National
Foundation, the Russian Foundation for Basic Research (08-02-00767-a),
and the program `Quantum Macrophysics' of the RAS.
\end{theacknowledgments}

\end{document}